# The Formation of Trust in Autonomous Vehicles after Interacting with Robotaxis on Public Roads

Xiang Chang[1,2,#], Zhijie Yi[1,#], Yichang Liu[1], Hongling Sheng[1], Dengbo He[1,*]

[1] Thrust of Intelligent Transportation, The Hong Kong University of Science and Technology (Guangzhou), Guangdong, China

[2] Cornell tech, Cornell University, New York, USA

# indicates co-first author

This study investigates how pedestrian trust, receptivity, and behavior evolve during interactions with Level-4 autonomous vehicles (AVs) at uncontrolled urban intersections in a naturalistic setting. While public acceptance is critical for AV adoption, most prior studies relied on simplified simulations or field tests. We conducted a real-world experiment in a commercial Robotaxi operation zone, where 33 participants repeatedly crossed an uncontrolled intersection with frequent Level-4 Robotaxi traffic. Participants completed the Pedestrian Behavior Questionnaire (PBQ), Pedestrian Receptivity Questionnaire for Fully AVs (PRQF), pre- and post-experiment Trust in AVs Scale, and Personal Innovativeness Scale (PIS). Results showed that trust in AVs significantly increased post-experiment, with the increase positively associated with the Interaction component of PRQF. Additionally, both the Positive and Error subscales of the PBQ significantly influenced trust change. This study reveals how trust forms in real-world pedestrian-AV encounters, offering insights beyond lab-based research by accounting for population heterogeneity.

## INTRODUCTION

The safe and efficient operation of autonomous vehicles (AVs) relies on more than technological advancement. The social acceptance of the AVs can also affect how users adopt and interact with the AVs (Velasco et al., 2021; Winkle et al., 2016; Zhou et al., 2021), which is especially the case in mixed traffic consisting of AVs and vulnerable road users, such as cyclists and pedestrians (Hancock et al., 2019; Rasouli et al., 2019). Previous research found that pedestrians may hesitate when interacting with AVs at uncontrolled intersections (de Miguel et al., 2019; Hochman et al., 2020; Rasouli et al., 2017), potentially leading to "Freezing Robot Problem" (Trautman et al., 2010) where pedestrians hinder the AVs from making any movements. Such a phenomenon may be caused by road users' mis-calibrated trust in and perception of AVs.

Thus, previous research has investigated how pedestrians interact with AVs and the factors influencing their attitudes towards AVs. However, on one hand, beyond age, gender, and cultural background, the effects of various individual characteristics, such as personality traits and personal innovativeness (PI), on pedestrians' trust in AVs have been under-investigated. Existing research regarding users' trust in automation has identified associations between trust and users' personality dimensions, including neuroticism and extraversion (Merritt et al., 2008; Hoff et al., 2015). Additionally, PI, which reflects an individual's tendency to accept novel experiences and technologies, has been found to be associated with openness to AV adoption (Deb et al., 2017).

On the other hand, existing research on pedestrian-AV interactions primarily relied on laboratory simulations or closed-road field studies, which, though, has provided valuable theoretical insights (Chang et al., 2024; Clamann et al., 2017; Mahadevan et al., 2019), often oversimplified the complexity of real-world traffic environments by neglecting the dynamic and unpredictable characteristics of urban ecosystems (Beggiato et al., 2017) and nullified the risks that pedestrians may face in real-world traffic scenarios. As a result, there is still a gap regarding how pedestrians' trust in AVs evolves in dynamic, highly interactive real-world scenarios. Addressing this gap is essential for accurately capturing the evolution of pedestrians' attitudes toward AV technology and providing actionable insights for optimizing AV design.

Thus, a user experiment was conducted at a real-world uncontrolled urban road intersection where participants interacted with commercially running robot taxis that can be categorized as Level 4 by the Society of Automotive Engineers (SAE) (SAE international, 2021). By evaluating participants' perception of AVs before and after they interacted with the AVs, for the first time, we explored how interacting with the AVs can affect the formation of participants' trust in AVs. Further, given that individual differences, such as personality (Kraus et al., 2021; Nordhoff et al., 2025) and personal innovativeness (Hegner et al., 2019) have been found to influence users' trust in AVs, we also explored the factors that can moderate pedestrians' trust in AVs during the interaction process.

## METHODS

### Experimental Site

As shown in Figure 1, the experiment was conducted at an uncontrolled intersection with a zebra crossing but without a traffic light on a straight, two-way urban road. The zebra crossing spanned approximately 15 meters. This intersection lies within an operating zone of commercial Robotaxis with a mix of road users, including AVs, human-driven vehicles, buses, trucks, and non-motorized vehicles. In addition to the commercially operating AVs that randomly travelled across the intersection, during the experiment, an SAE L4 AV provided by Pony AI also travelled repeatedly across the intersection, where there was a safe driver in the front seat, but he never intervened the AV during the experiment. All AVs, including the one provided by Pony AI, were equipped with a rooftop light that would clearly indicate that they were operating in autonomous driving mode.

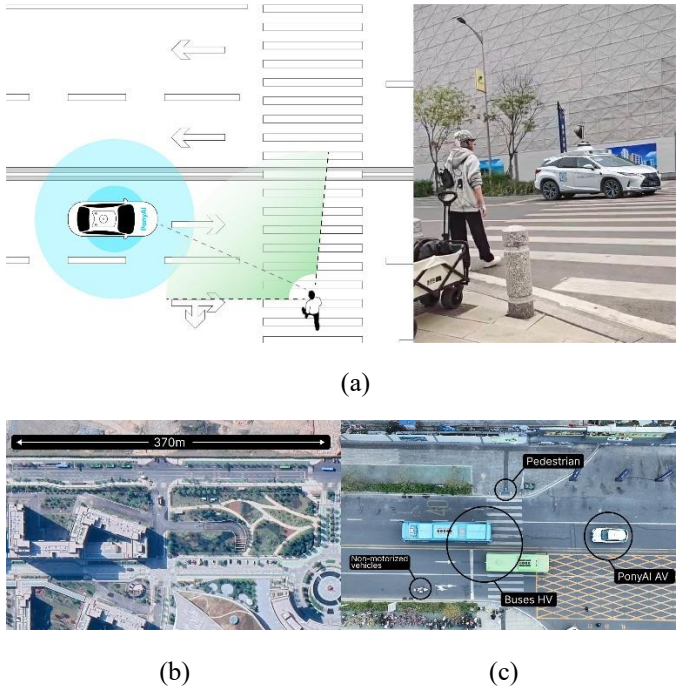

(a)

(b) (c)

Figure 1. Layout of the experiment site: a) Schematic diagram and photos of pedestrian-autonomous vehicle interaction in the experiment; b) A satellite image of the entire experimental site; c) A drone image showing the complex traffic conditions during the experiment.

**Participants**

In total, 33 participants participated in this experiment, including 19 males and 14 females. Among them, 15 participants were aged from 18 to 24 years old, 16 were aged from 25 to 34 years old, and 2 were over 55 years old. All participants lived in the nearby regions of the experimental site and had normal mobility and vision (or corrected vision). Prior to the experiment, participants received an informed consent form, and each participant was compensated 120 RMB (roughly USD 16.5) after completing the study.

**Procedure**

Before the day of the experiment, participants were asked to complete demographic questionnaires, a 4-item Personal Innovativeness Scale (PIS) adapted from Agarwal and Prasad (Agarwal et al., 1998), measuring their openness to new technologies, and a 20-item Pedestrian Behavior Questionnaire (PBQ) (Deb et al., 2017), measuring their risky pedestrian behavior.

After arriving at the experiment site, the experimenter introduced the Robotaxi by Pony AI to the participants. The participants were informed that it is a fully unmanned self-driving car, operating in autonomous driving mode, and pedestrians need to cross the intersection as many times as possible based on their own judgment. Then, they completed a pre-experiment Trust Scale (Manchon et al., 2022; Wang et al., 2024) to evaluate their initial trust in AVs. Next, participants were put on sensors to collect their physiological responses and record their behavioral performance during the interaction with the AV. Once ready, the Robotaxi by Pony AI started to drive. At the same time, one minute of resting data from the participant was collected, during which they stood beside the intersection, maintained steady breathing, and looked straight ahead. The analysis of the physiological responses is out of the scope of this study, due to the page limit, though the data collection process is reported here to avoid misleading.

Then, the formal experiment started, which lasted 20 minutes, during which participants were required to repeatedly cross the crosswalk while ensuring safety. During each cross, participants needed to observe and judge the intentions of other vehicles, including the AV, before they felt safe to cross the road. The experimental scene was recorded using a drone flying at 50 meters (see Figure 1c), capturing the trajectories of pedestrians and all other road agents during the interaction.

After the experiment, participants were allowed to rest and required to complete a post-experiment questionnaire, including the Trust Scale measuring their trust in AVs, and a 16-items Pedestrian Receptivity Questionnaire for fully AVs (PRQF) (Deb et al., 2017) measuring their receptivity to AVs as a pedestrian.

**Variables and Statistical Models**

The PIS was a 7-point Likert scale, with a higher overall score indicating greater personal innovativeness (Agarwal et al., 1998). The PRQF is a 7-point Likert scale (1 = strongly disagree to 7 = strongly agree) measuring three subscales: Safety, Interaction, and Compatibility (Deb et al., 2017; Dommes et al., 2024). Our study adopted mean scores for each subscale as independent variables, excluding one question ("*Interacting with the system would not require a lot of mental effort*"), to ensure consistency with previous research. The PBQ is a 6-point scale (Deb et al., 2017), and the questions targeted five subscales (i.e., Violation, Error, Lapse, Aggressive and Positive), with higher scores representing riskier pedestrian behavior. The Trust Scale is a 6-point questionnaire adapted from prior studies that measures trust in automation (Manchon et al., 2022; Wang et al., 2024), with higher mean score of all questions indicating higher trust in AVs.

Before fitting the model, we assessed the correlation between all independent variables to account for potential multicollinearity issues. Then, to explore how trust in AVs changed after interacting with AVs, a mixed linear effects model was built, with trust in AV as the dependent variable and the timing (pre-experiment vs. post-experiment), PIS score, three subscales of PRQF, five subscales of PBQ, and their two-way interactions with the timing as independent variables in the full model. The model was constructed in SAS OnDemand using Proc Mixed. Repeated measures on participants were accounted for using the Generalized Estimation Equation (GEE). Backward model selection was conducted based on the Bayesian Information Criterion (BIC) (Lee et al., 2014).

## RESULTS

We report all significant effects (*p* < .05) in the final model after model selection. Note that all variables removed during the model selection process were non-significant (*p* > .05).

Table 1. Results for mixed model fixed effects.

| Effect | β | SE | F value | *p*-value |
| --- | --- | --- | --- | --- |
| Timing | -0.56 | 0.43 | F(1,31) = 1.74 | .2 |
| Interaction | 0.69 | 0.09 | F(1,29) = 65.29 | <.001* |
| Interaction * Timing | -0.31 | 0.34 | F(1,31) = 7.26 | .01* |
| Error | 0.09 | 0.04 | F(1,29) = 5.99 | .02* |
| Positive | 0.05 | 0.02 | F(1,29) = 5.97 | .02* |

Note: β is the estimated effect, SE is the standard error, * marks a significant (*p* < .05) effect.

First, as shown in Table 1 and visualized in Figure 2, we observed the main effects of Error and Positive subscales from the PBQ on pedestrians' trust in AVs. Specifically, for every 1-unit increase in Error (e.g., tendency to wrongly walk on cycling paths and crossing the street without observing traffic), the trust increased by 0.09, with a 95% confidence interval (95%CI) ranging from 0.02 to 0.2, indicating that participants who frequently reported inattentive or unintentional behaviors also tended to report higher trust in AVs. At the same time, every 1-unit increase in Positive component (e.g., cautious crossing and politely yielding) was associated with a 0.05 decrease in trust in AVs, 95%CI: [0.01 to 0.09]. Given that the Positive is reversely scaled in the PBQ (Deb et al., 2017), this positive association indicates that pedestrians with less positive behavior (e.g., lower tendency to yield to vehicles) in traffic environments tended to trust more in AVs.

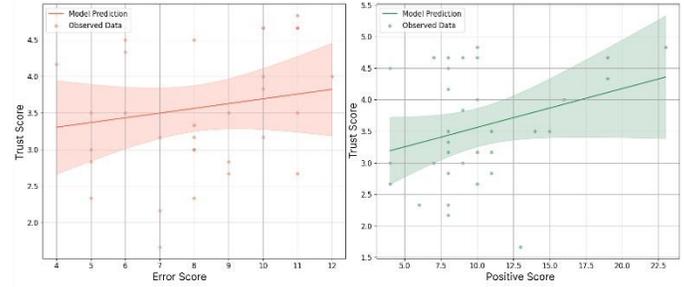

Figure 2. Visualization of the marginal effects of the Error and Positive effects.

Further, as visualized in Figure 3, we observed a two-way interaction effect between the Timing and the Interaction component in PRQF (e.g., their perceived safety when crossing in front of AVs). In general, participants' trust increased after interacting with the AVs, and participants reported higher Interaction scores also reported higher trust in AVs, but the differences were not uniform across participants. Specifically, for every 1 unit increase in a participant's Interaction score, the pre-experiment trust and post-experiment trust increased by 0.69 (t(29) = 8.08, *p* < .0001, 95%CI: [0.52, 0.87]) and 0.82 (t(45) = 8.39, *p* < .0001, 95%CI: [0.62, 1.02]), respectively.

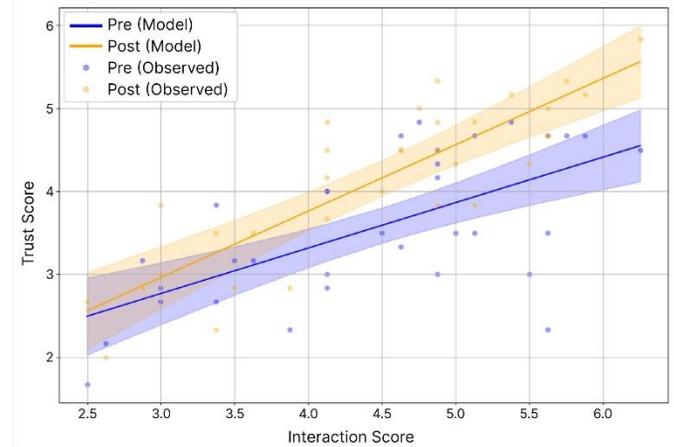

Figure 3. Visualization of the marginal effects of the time * interaction effect.

## DISCUSSION

In an on-road study, for the first time, we evaluated how interacting with AVs would affect pedestrians' trust in AVs. We found that certain characteristics of the pedestrians would moderate their formation of trust in AVs.

First, we found that lower Positive scores and higher error-prone behaviors were both associated with increased trust in AVs. Participants with higher Error scores are more likely to engage in risky street crossing behaviors (e.g., conducting inattentive street crossing or misjudging the gaps). When interacting with human-driven vehicles, such errors may lead to conflicts or near-crashes on the road. In our experiment, however, when the pedestrians exhibited errors, the AVs would always respond safely. This good performance of AVs would reinforce participants' impression of AV reliability and thus increase their trust in AVs. This may also explain why the pedestrians who got higher scores in the subscale of "Positive" (lower tendency to show positive behaviors when interacting with vehicles) also trusted more in AVs. Specifically, those who acted less positively were also more likely to receive more positive feedback from the AVs and thus trusted the AVs more, especially after interacting with them. Conversely, pedestrians who typically yielded even when they had the right-of-way may have continued to do so when encountering AVs, resulting in few trust-boosting interactions and thus had a slower trust increase.

Further, similar to what has been found among passengers of AVs (Wang et al., 2024; Xie et al., 2025), we observed an increase in trust after participants interacted with the AVs, demonstrating that the interaction between pedestrians and AVs can influence pedestrians' trust in AVs. This finding aligns with prior research (Muir et al., 1996), which suggests that trust is often cultivated through direct interaction with a system (if the system performed well during the interaction), supporting the notion that "*trust is built progressively through interaction*." In our experiment, as mentioned previously, the AVs consistently exhibited safe and predictable behaviors (e.g., slowing down or stopping when pedestrians approached), and thus, the pedestrians' positive impression of the system may have been gradually strengthened.

Further, we also found that the influence of interacting with the AVs can be moderated by the pedestrians' Interaction score, with the post-experiment trust being more sensitive to the Interaction score than the pre-experiment score. This finding agrees with the trust model by Lee et al. (Lee et al., 2004), which indicated that variations in trust can be affected by factors in the dispositional layer (e.g., tendency to trust the technologies in general) and the learned trust during the interaction with the system. Specifically, as the participants had not yet experienced the AV before the experiment, the pre-experiment trust was moderated only by their prior impression or knowledge of the AV. After interacting with the automation and when the performance of the automation matched their prior impression, the influence of the prior impression might be enhanced by their actual experience with the system.

Finally, it should be noted that our experiment still has limitations. First, given that this is an experiment on public roads, we were not able to control the density of other road agents, including other pedestrians and human-driven vehicles, which might affect participants' behaviors and perceptions of the AVs. Second, although the participants interacted with AVs freely, they were still required to put on several data collection sensors, and they were intentionally told to cross the streets as many times as possible. This experiment setup may also bias the participants' behaviors to some level, though we kept real-world risks that are difficult to replicate in simulated environments.

Finally, to ensure safety, the AVs in our experiment all prioritized safety. Future research should also explore how failures or near-crashes may affect public acceptance of AVs.

## CONCLUSIONS

Unlike simulation studies or field studies, for the first time, we allowed participants to interact with the AVs in a real-world environment with commercially running AVs. We found that with reliable AV technologies, pedestrians' trust would increase after interacting with the AVs. Further, personality factors can moderate the formation of trust in AVs. Such findings indicate that future AV providers should consider the heterogeneity of the public. Future research should also explore ways to increase trust and interaction efficiency in real-world settings, for example, by deploying external human-machine interfaces on the AVs.

## CONFLICT OF INTEREST

The authors declare no potential conflicts of interest with respect to the research, authorship, and/or publication of this article.

## AKNOWLEDGEMENTS

This work is supported by Nansha District Key Area S&T Scheme, Science and Technology Bureau of Nansha District (No. 2023ZD006), and in part by the Guangzhou Municipal Science and Technology Project (No. 2023A03J0011) and Guangdong Provincial Key Lab of Integrated Communication, Sensing and Computation for Ubiquitous Internet of Things (No. 2023B1212010007). This work was assisted by Pony.ai in vehicle dispatching.